\documentclass{article}

\usepackage{arxiv}

\usepackage[utf8]{inputenc} 
\usepackage[T1]{fontenc}    
\usepackage{hyperref}       
\usepackage{url}            
\usepackage{booktabs}       
\usepackage{amsfonts}       
\usepackage{nicefrac}       
\usepackage{microtype}      
\usepackage{lipsum}		
\usepackage{graphicx}
\usepackage{natbib}
\usepackage{doi}
\usepackage{amssymb}
\usepackage{lineno}
\usepackage{color}
\usepackage{wrapfig}
\usepackage{amsmath}
\usepackage[dvipsnames]{xcolor}
 \usepackage{adjustbox}
\usepackage{subcaption}
\usepackage{tabularx}
\usepackage{float}
\usepackage{hyperref}
\usepackage{url}
\usepackage{listings}

\usepackage{color}
\usepackage{caption}
\usepackage{subcaption}
\definecolor{codegreen}{rgb}{0,0.6,0}
\definecolor{codegray}{rgb}{0.5,0.5,0.5}
\definecolor{codepurple}{rgb}{0.58,0,0.82}
\definecolor{backcolour}{rgb}{0.95,0.95,0.92}

\lstdefinestyle{mystyle}{
    backgroundcolor=\color{backcolour},
    commentstyle=\color{codegreen},
    basicstyle=\ttfamily\scriptsize
}

\title{A modelling framework for the analysis of the SARS-CoV2 transmission dynamics}

\author{Anastasia Chatzilena\\
	Department of Engineering Mathematics\\
	University of Bristol\\
	\texttt{a.chatzilena@bristol.ac.uk} \\
	\And
	Nikolaos Demiris\\
	Department of Statistics\\
	Athens University of Economics and Business\\
	\texttt{nikos@aueb.gr} \\
	\AND
	Konstantinos Kalogeropoulos \\
	Department of Statistics\\
    London School of Economics and Political Science\\
	\texttt{k.kalogeropoulos@lse.ac.uk} \\
}

\renewcommand{\shorttitle}{Modelling SARS-CoV2 transmission}

\hypersetup{
pdftitle={A modelling framework for the analysis of the transmission of SARS-CoV2},
pdfsubject={stat.AP},
pdfauthor={Anastasia Chatzilena, Nikolaos Demiris, Konstantinos Kalogeropoulos},
}

\begin{document}

\maketitle

\begin{abstract}
Despite the progress in medical data collection the actual burden of SARS-CoV-2 remains unknown due to under-ascertainment of cases. This was apparent in the acute phase of the pandemic and the use of reported  deaths has been pointed out as a more reliable source of information, likely less prone to under-reporting. Since daily deaths occur from past infections weighted by their probability of death, one may infer the total number of infections accounting for their age distribution, using the data on reported deaths. We adopt this framework and assume that the dynamics generating the total number of infections can be described by a continuous time transmission model expressed through a system of non-linear ordinary differential equations where the transmission rate is modelled as a diffusion process allowing to reveal both the effect of control strategies and the changes in individuals behavior. We develop this flexible Bayesian tool in Stan and study 3 pairs of European countries, estimating the time-varying reproduction number($R_t$) as well as the true cumulative number of infected individuals. As we estimate the true number of infections we offer a more accurate estimate of $R_t$. We also provide an estimate of the daily reporting ratio and discuss the effects of changes in mobility and testing on the inferred quantities.
\end{abstract}


\section{Introduction}
\label{Section1}

The SARS-CoV-2 pandemic which originated in December 2019 in the city of Wuhan, China and spread rapidly across the globe, has had devastating economic and social consequences, in addition to the severe loss of human life. Early on the pandemic, there have been consistent efforts from national health authorities to publicly report the daily counts of laboratory-confirmed cases and deaths in real-time, however, it became more than apparent that surveillance was going to be a challenging part in our quantitative understanding due to the low diagnostic capacity of many asymptomatic and mild infections. In response to the rising numbers of reported cases and deaths, which characterized the pandemic waves in early and late 2020 and mid-2021, many European countries, have implemented several control strategies, ranging from social distancing recommendations to large-scale lockdowns. Aiming to reduce a key epidemiological parameter, the time-varying reproduction number, these control strategies have been changing over time and different countries have adopted different action plans, expressing different policy decisions, social mechanisms, health systems capacity and transmission dynamics.  

The evaluation of the effectiveness of the implemented preventive measures in each country, as reflected in the reduction of transmission, is not straightforward. Inference from infectious disease data is a non-standard problem since one rarely observes the necessary evidence \citep{rhodes1996counting,demiris2005bayesianRSS}. Despite the massive progress in data collection, this has become apparent in the current pandemic in the estimation of the time-varying reproduction number that typically relies on surveillance data which are usually biased and incomplete. It is hard to evaluate the actual burden of the disease when we deal with such a highly transmissible disease as COVID-19, with many asymptomatic and mild symptomatic infections which are not detected by health systems~\citep[]{jombart2020inferring, li2020substantial,verity2020estimates}. Some large-scale seroprevalence studies~\citep[]{ward2021sars, pollan2020prevalence} aimed to estimate the actual number of infections and found severe under-ascertainment. Depending on the testing capacities imposed by healthcare resource constraints and the adopted testing and tracing policies, the level of under-ascertainment has been changing over time and across countries.

The number of reported deaths is a more reliable indication of which countries around the globe have faced the most severe effects of the SARS-CoV-2 pandemic and even though, the reporting of deaths may vary over time and across countries, data on reported deaths are likely less prone to under-reporting. Therefore, given that daily deaths occur from past infections weighted by their probability of death, we can infer the total number of infections using the data on reported deaths~\citep[]{jombart2020inferring, flaxman2020estimating}. 

This work uses a model-based approach to estimate the transmissibility of SARS-CoV-2 and the effect of the adopted control measures across 6 European countries. We introduce an extension of the \citet{dureau2013capturing} model where we fit to the unknown true number of cases an SEIR (susceptible-exposed-infected-recovered) compartmental model driven by a stochastic time-varying transmission rate that captures the effect of both the control measures and the behavioural changes. Our model may also be viewed as an extension of the work in ~\citet[]{flaxman2020estimating} where the estimated number of cases is indirectly inferred and generated by a diffusion-driven stochastic SEIR process.

 We implement our suggested model in an evidence synthesis Bayesian framework incorporating three different sources of information; reported deaths, reported cases and individual-case data. We use Hamiltonian Monte Carlo employing the Stan software, to fit our model to daily reported deaths for Greece, Portugal, United Kingdom, Germany, Sweden and Norway. The proposed Bayesian tool can be adapted to other countries in a straightforward manner. We then examine how we can combine our estimates of the total number of daily cases with data on daily laboratory-confirmed cases and estimate the daily reporting ratio. Finally, through a multivariate regression analysis, we disentangle the effects of preventive measures and testing policies on the estimated total cases, the time-varying transmission rate and the reporting rate, using only publicly available data. Our proposed model is then fit to data from country pairs with similar population demographics and health and social welfare infrastructures, to gain insights on the COVID-19 pandemic.

The paper is organised as follows: In Section \ref{Section2} we present the developed model and adopted methods. Specifically, we provide an overview of the available data and present analytically our modelling framework. Section \ref{Section3} contains the results of our empirical analysis. Section \ref{Section4} concludes and provides some relevant discussion. All the data, R and Stan code files are made freely available at  \url{https://github.com/anastasiachtz/seir-gbm.git}.

\section{Methods}
\label{Section2}

\subsection{The data to date}
\label{Section2.1}

Publicly available datasets containing surveillance data on new confirmed cases and deaths per day and per country or region, are maintained in the COVID-19 Data Repository by the Center for Systems Science and Engineering (CSSE) at Johns Hopkins University (JHU)~\cite[]{dong2020interactive} based on various sources. To estimate our model parameters describing the mechanisms of disease spread we use only data on the reported number of deaths as a more reliable source of information compared to laboratory-confirmed cases. 

The number of laboratory-confirmed daily cases constitutes a biased source of information, primarily due to the high proportion~\cite[]{lavezzo2020suppression,jombart2020inferring, li2020substantial,verity2020estimates, russell2020reconstructing} of mild or asymptomatic infections which are not typically reported since people may not seek medical care or testing. This phenomenon intensifies during the pandemic waves, when the health care systems are overwhelmed, people with mild COVID-19 symptoms are advised to avoid health care unless it is necessary. Taking also into account the testing capacity constraints of each country, which has been changing over time, casts further doubts regarding the quality of the daily cases, as a source of data, towards estimating the actual number of infected people during the pandemic. The latter has been confirmed by some large-scale seroprevalence studies~\cite[]{ward2021sars, pollan2020prevalence}. In this regard, we consider the total number of infections to be unobserved (latent).
 
Reported deaths are offering a more reliable quantitative understanding of the pandemic despite their own limitations. \citet[]{flaxman2020estimating} suggested calculating backwards from observed deaths to the number of infections. The number of deaths attributed to COVID-19 has been considered less prone to under-reporting since deaths mainly arise from severe cases who are more likely to have been tested while seeking health care or after death. Therefore, we assume that the number of unreported deaths in the countries under study is negligible. However, early in the pandemic, in the absence of European and international standards, some deaths due to COVID-19 may have not be recorded. We begin our analysis 2 weeks prior each country reports 10 cumulative deaths following \citet[]{flaxman2020estimating}, with knowledge of the incomplete nature of these early data since our estimates are not affected by them. We are aware that the timing of the reporting procedures may differ between countries and reporting delays may exist, but we consider that they are relatively minor for the countries under study, however, we incorporate uncertainty within our observational model. In our context, deaths offer a window to the past, revealing the infections which led to them, but they can not be used in real-time analysis to inform current infections unless additional assumptions are imposed. 
 
\subsection{Modelling framework}
\label{Section2.2}

\begin{figure}
    \centering
    \caption{Directed  acyclic  graph  of the model. Square nodes represent observable quantities, black circles are latent quantities and orange circles are latent quantities informed by data. Solid arrows represent stochastic dependencies and dashed arrows represent deterministic dependencies.\\
    $^*$~\citet[]{levin2020assessing}, $^{**}$~\citet[]{li2020early,liu2020time}, $^{\#}$~\citet[]{verity2020estimates}
    }
    \includegraphics[width=\linewidth]{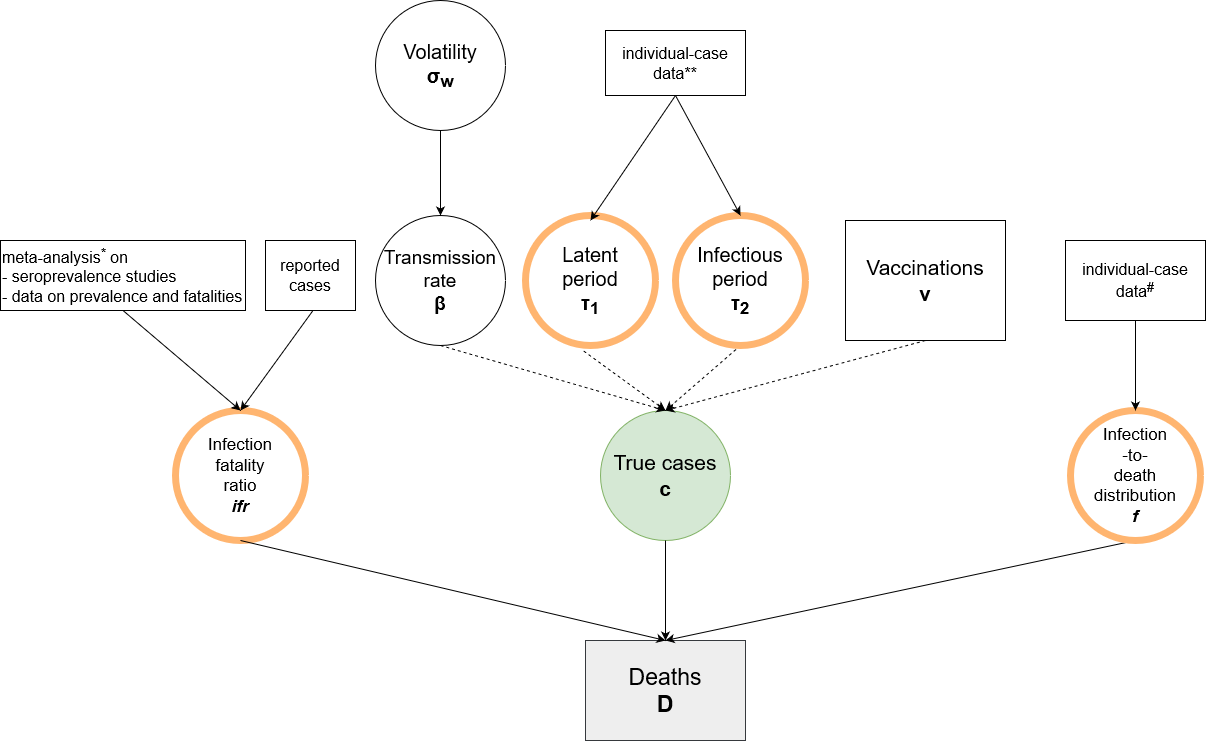}
    \label{fig:fullDAG}
\end{figure}

Based only on publicly available data sources on the reported number of deaths, we develop a modelling framework where we link the daily new infections to the reported deaths, and infections are generated by a stochastic transmission SEIR compartmental model, adding another level of hierarchy to the model. Also, given the significant vaccine rollout, we extend our model to reflect the impact of vaccinations. The directed acyclic graph at figure \ref{fig:fullDAG} represents the structure of the model delineating its transmission and observational components. Statistical inference procedures are addressed within a Bayesian framework and parameter learning is carried out using Stan's implementation of Hamiltonian Monte Carlo~\cite[]{Stan2018Stan}. The analysis period ranges from March 2020 to September 2021. Death count data for Greece, Portugal, Germany, United Kingdom and Norway are obtained from CSSE at JHU~\cite[]{dong2020interactive}, while data for Sweden were directly obtained from \href{https://www.folkhalsomyndigheten.se/smittskydd-beredskap/utbrott/aktuella-utbrott/covid-19/statistik-och-analyser/bekraftade-fall-i-sverige/}{Folkhälsomyndigheten}, the Public Health Agency of Sweden. Swedish Public Health Agency adjusts the daily number of deaths ex-post, correcting for the reporting delay, resulting in significantly different counts compared to their initial reports. We examine all possible discrepancies between data reported by national health authorities and data maintained in the COVID-19 Data Repository by CSSE at JHU and use the most integrated dataset for each country. Detailed references for each data source are presented in the supplementary material. 

\subsection*{Observation Model}
\label{observ_model}

The base of the model is to link the daily new infections to the data on reported deaths. The infection fatality ratio serves as a bridge between deaths and true infections, in the sense that deaths on any day occur from previously acquired infections according to their probability of death given infection. Following \citet[]{flaxman2020estimating}, we assume that the expected number of deaths at time $t$, $d_t$, is a function of the unobserved true past infections $c_{t-s}$, weighted by the distribution of time from infection to death, $f$ ~\cite[]{verity2020estimates}, and multiplied by their probability of death i.e the infection fatality ratio, $ifr$~\cite[]{levin2020assessing}. Therefore, the expected number of deaths at time $t$ can be expressed as,
 \begin{equation}
    {d}_{t} = ifr_t * \sum_{\tau=1}^{t-1} c_{\tau} f_{t-\tau}
\end{equation}
where $c_{\tau}$ are  the  unobserved  true  past  infections which will be linked to the solution of a system of ODEs \eqref{eq:ct}. The infection-to-death distribution is the sum of two independent Gamma distributions, the estimated infection-to-onset distribution and the estimated onset-to-death distribution~\cite[]{verity2020estimates, flaxman2020estimating}, thus resulting in the $f \sim \Gamma(6.29, 0.26)$ distribution, which is discretized by $f_1 = \int_{0}^{1.5} f(\tau) d\tau$ and $f_s = \int_{s-0.5}^{s+0.5} f(\tau) d\tau$ for $s=2,3,...$. 

The overall $ifr$ depends on the age distribution of the infections, which changes over time. We consider that $ifr_t$ is a piecewise constant process that changes values in predetermined time periods which we specify by identifying the points in time where the age distribution of reported infections changes, especially for the $50-69$ and over $70$ age groups which have both the lower under-reporting rates and the higher age-specific $ifr$. We evaluate different overall $ifr$ across these different periods using estimates of the age-specific $ifr$ reported by the Centers for Disease Control and Prevention (CDC) which are based on a meta-analysis of seroprevalence studies and data on prevalence and fatalities from countries with extensive tracing programs~\cite[]{levin2020assessing}, capturing the ratio of fatalities to total infections. Then, we adjust these estimates taking into account the improvement of health infrastructures after the first few months of the pandemic and the emergence of more lethal and transmissible variants at the end of 2020~\cite[]{davies2021estimated,volz2021assessing}. Our central calculations of the overall $ifr$ for each time period in the year 2020, of a relatively constant proportion of reported cases per age group, are based on
\begin{equation}
 \overline{ifr_t} = \sum_{g=1}^{G}ifr^g \frac{c_{g}^{rep}}{C^{rep}} 
\end{equation}
where $ifr^g$ is the age-specific $ifr$ for age group $g$, $g=1,\dots,G$, $c_{g}^{rep}$ is the cumulative number of reported cases of age-group $g$ and $C^{rep}$ is the cumulative number of all reported cases. Given our calculations on mean $\overline{ifr_t}$s for every country, we assign $\mathrm{Beta}$ priors for each $ifr_t$ with mean $\overline{ifr_t}$ and low variance. Note that the scale of the estimated number of infection will depend upon the exact $ifr$ value but most of the other quantities we estimate, including $R_t$ are unlikely to be materially affected by moderate variations to $ifr$.

At the end of 2020 and the beginning of 2021, COVID-19 vaccines became available in several countries, including the countries under study. Vulnerable groups at the highest risk of severe disease were prioritized by the designed vaccination programs, therefore older age groups were immunised first. Protection of older age groups and an increase in the number of infections in younger age groups in the presence of more transmissible variants, led to a significant decrease in $ifr$. Based on the rate of decrease of the estimated overall $ifr$ in the UK by Birrell et al. ~\citep[]{birrell2021real}, we re-adjust the overall $ifr$ in all countries, taking into account the timing of their immunization programs. 

The reported deaths $D_{t}$ at time $t$ are assigned a Negative Binomial distribution with mean $d_t$ and variance $d_t + \frac{d^{2}_{t}}{\phi}$ denoted as
\begin{equation}
    D_{t} \sim \operatorname{Negative \hspace{0.1cm} Binomial} \left(d_t,d_t + \frac{d^{2}_{t}}{\phi}\right)
\end{equation}
where $1/\phi$ controls the overdispersion and a-prioti $1/\phi \sim C^{+}(0,5)$.

\subsection*{Transmission Model}
\label{transm_model}

Several epidemiological models have been proposed~\cite[]{birrell2021real,flaxman2020estimating,lemaitre2020assessing,dehning2020inferring,wood2021inferring,kucharski2020early,hao2020reconstruction} attempting to describe the transmission dynamics of SARS-CoV-2 and the effects of preventative measures on these dynamics. Non-pharmaceutical interventions such as social distancing recommendations, limitations on the size of indoor and outdoor gatherings, promotion of teleworking, self-isolation of symptomatic individuals, school closures and ultimately stay-at-home measures, primarily aim to limit the contact rate between individuals while also affecting the relative infectiousness of infected individuals. These control strategies have been changing over time, different action plans have been adopted according to the epidemiological situation of each country and local communities have responded differently to these measures between pandemic waves. These considerations suggest that a flexible model should be adopted for capturing the dynamics of the effective transmission rate.

Here we consider a stochastic expansion of the well-known deterministic SEIR compartmental model~\cite[]{anderson1992infectious} which assumes a homogeneously mixing population in which all individuals are equally susceptible and equally infectious should they become infected. Instead of a constant transmission rate between susceptible and infected individuals, we assume that it follows a stochastic process~\cite[]{dureau2013capturing}. Specifically, conditional upon the stochastic infection rate $\beta_t$, we consider that the transmission dynamics resulting to the generated true infections in each country, are expressed as the solution of the following system of non-linear ordinary differential equations (ODEs):
\begin{align}\begin{split}
\frac{\mathrm{d}S_t}{\mathrm{d}t} & = - \beta_t S_t \frac{\left(I_{1t}+I_{2t}\right)}{N} - \rho \nu_{t-U}\\
\frac{\mathrm{d}E_{1t}}{\mathrm{d}t} & =  \beta_t S_t \frac{\left(I_{1t}+I_{2t}\right)}{N} - \gamma_1 E_{1t}\\
\frac{\mathrm{d}E_{2t}}{\mathrm{d}t} & = \gamma_1 E_{1t} - \gamma_1 E_{2t}\\
\frac{\mathrm{d}I_{1t}}{\mathrm{d}t} & = \gamma_1 E_{2t} - \gamma_2 I_{1t}\\
\frac{\mathrm{d}I_{2t}}{\mathrm{d}t} & = \gamma_2 I_{1t} - \gamma_2 I_{2t}\\
\frac{\mathrm{d}R_{t}}{\mathrm{d}t} & =  \gamma_2 I_{2t} +\rho \nu_{t-U}
\label{eq:ode_vac}
\end{split}\end{align}
where $S_{t}$ represents the number of susceptible, $E_{t}$ the number of exposed, but not yet infectious, $I_{t}$ the number of infected and $R_{t}$ the number of recovered individuals at time t. The total population size of each country is denoted by $N$ (with $N = S_{t} + E_{t} + I_{t} + R_{t}$). In order to allow the latent and infectious periods to be gamma distributed, we assume that each of the E and I compartments are defined by two classes, $E_{1t}$, $E_{2t}$ and $I_{1t}$, $I_{2t}$ respectively. Hence, $\gamma_1$ denotes the rate at which the exposed individuals become infective so that $\frac{2}{\gamma_1}$ is the mean latent period and $\gamma_2$ denotes the recovery rate so that $\frac{2}{\gamma_2}$ is the mean infectious period.

Employing an SVEIR(susceptible–vaccinated–exposed–infectious–recovered) model is out of the scope of this paper, however, we consider a simple vaccination scenario, where vaccinated individuals are protected after $U=45$ days of receiving the first dose of any of the available vaccines, which is the average time to obtain immunity given that during this time interval they also have their second dose where necessary~\cite[]{polack2020safety}. Thus, $\nu_{t-U}$ is the reported number of individuals who received the first dose of a vaccine $U$ days prior to time t. To account for imperfect vaccine efficacy, we consider that vaccinated individuals move to the removed compartment proportionally to the vaccines' efficacy which is denoted by $\rho$ and we set it equal to $50\%$ as an average efficacy of the different types of the distributed vaccines.

The transmission rate at time t is denoted by $\beta_{t}$, for which we assume the following stochastic differential equation (SDE)

\begin{align}
\begin{split}
\label{eq:beta_prior}
\mathrm{d}\eta_t & = \mu(\eta_t,\theta_{\eta}) + \sigma(\eta_t,\theta_{\eta})\mathrm{d}B_t\\
\eta_{t} & =g(\beta_{t}).
\end{split}\end{align}

The model in \eqref{eq:beta_prior} may be viewed as the prior for the transmission rate trajectory $\beta_t$. The function $g(\cdot)$ transforms to the real line and is typically set to the logarithm $log(\cdot)$. The drift function $\mu(\cdot)$ determines the mean change in $\beta_t$ and is being set to $0$ as we a-priori assume that upward and downward movements are equally likely. The function $\sigma(\eta_t,\theta_{\eta})$ reflects the volatility and $B_t$ denotes standard Brownian motion. Our starting point is a constant volatility assumption, $\sigma(\cdot)\equiv \sigma$, but this is relaxed by introducing specific change-points across different waves to capture potential different responses resulting from adaptive human behavior. Given these specifications, we get the geometric Brownian motion as the prior for the $\beta_t$ trajectory.

In order to link with the available observations, the model-implied daily new infections, denoted by $c_t$, are obtained by
\begin{equation}
\label{eq:ct}
    c_t = \int_{t-1}^{t} \gamma_1 E_{2s} \mathrm{d}s.
\end{equation}
The above integral requires solving the ODE in \eqref{eq:ode_vac} together with the SDE in \eqref{eq:beta_prior} and in this respect, our model can be viewed as a hypo-elliptic diffusion. As such, it cannot be solved analytically meaning that the exact likelihood function for the observed data is intractable. For this reason, we adopt the data augmentation framework of \citet{dureau2013capturing}, in the spirit of \citet{roberts2001augmentation}, that employs time-discretization via the Euler approximation. The fineness of the discretization can be chosen by the user to control the approximation error. For illustration purposes, let us consider $\beta_t$ to be constant between each pair of days, i.e. for each $[t-1,t)$, noting that smaller intervals can also be used. The model in \eqref{eq:beta_prior} then implies that $\eta_{t}|\eta_{t-1}\sim \mathcal{N} \left({\eta_{t-1},\sigma^2}\right)$ and the solution of the ODEs in \eqref{eq:ode_vac} can be approximated using the trapezoidal rule. More sophisticated Runge-Kutta methods may also be used at the expense of higher computational cost.

\subsection*{Prior specification}
To complete the model specification, we consider Gamma prior distributions for the rate of loss of latency and the recovery rate, with small variances, reflecting 2 days average latent period and 4-5 days average infectious period~\cite[]{li2020early,liu2020time}. Due to the fact that the testing capacity of the countries under study has been scaled up, particularly during 2021, we assume that new cases are isolated slightly faster, translated in shorter infectious period. Therefore, we consider an average infectious period of 5 days during the first year of the pandemic and adopt a shorter average infectious period of 4 days for the last several months. Finally, a half-Cauchy prior is assigned for the volatility of the Brownian motion, $\sigma_w \sim C^{+}(0,5)$ for each pandemic wave $w$.

\subsection*{Computation}
The model was fitted in the Stan software using the NUTS algorithm. Inference for ODE-based models represents a non-trivial statistical problem. As the chosen MCMC algorithm explores the parameter space we are effectively solving an increasing number of ODE systems and the ODEs' behaviour varies for different parameter values~\cite[]{grinsztajn2020bayesian}. Therefore, sophisticated systems of ODEs can be computationally intensive. This became apparent when fitting our stochastic SEIR model for an extended time period. As the number of observations increased, so did the number of parameters given the time-varying nature of the transmission rate, resulting in relatively small effective sample sizes indicating slow mixing. To improve efficiency, we fit first the corresponding SIR model and then use the posterior estimates to uniformly draw initial values for the SEIR model. Details on the implementation can be found at the supplementary material.

\section{Results}
\label{Section3}

\subsection{Estimates of key epidemiological quantities}
\label{key_epi_par}

In order to assess the time course of the pandemic in the 6 European countries under study, estimates on the time-varying reproduction number, the number of daily infections and the cumulative infections per country are reported.

\begin{figure}
\captionsetup{format =plain}%
     \centering
         \begin{subfigure}{\linewidth}
             \centering
             \caption{\label{fig:R_GR}\textbf{Greece}}
             \includegraphics[width=\linewidth]{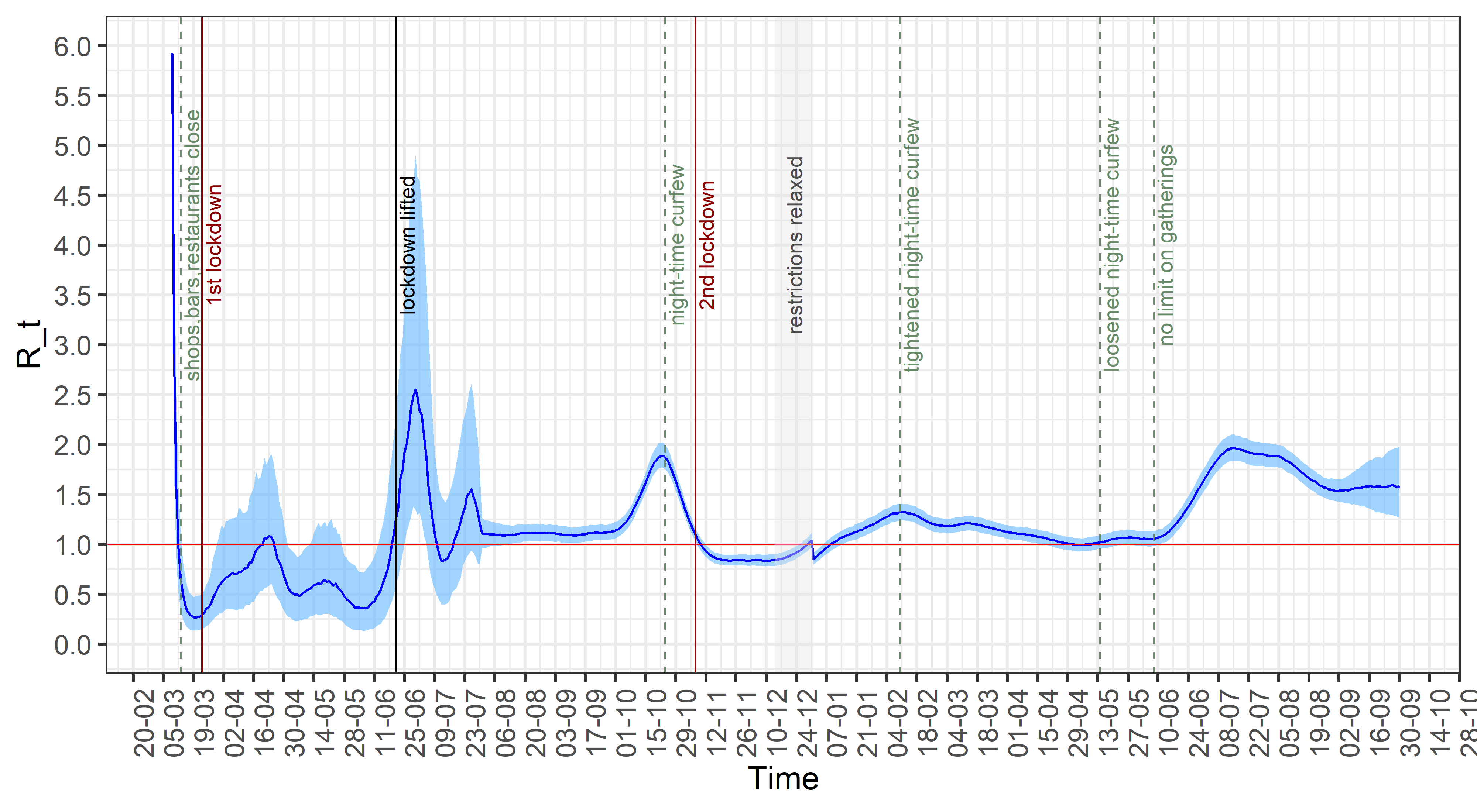}
         \end{subfigure}
         \par\bigskip
         \begin{subfigure}{\linewidth}
             \centering
             \caption{\label{fig:R_PT}\textbf{Portugal}}
             \includegraphics[width=\linewidth]{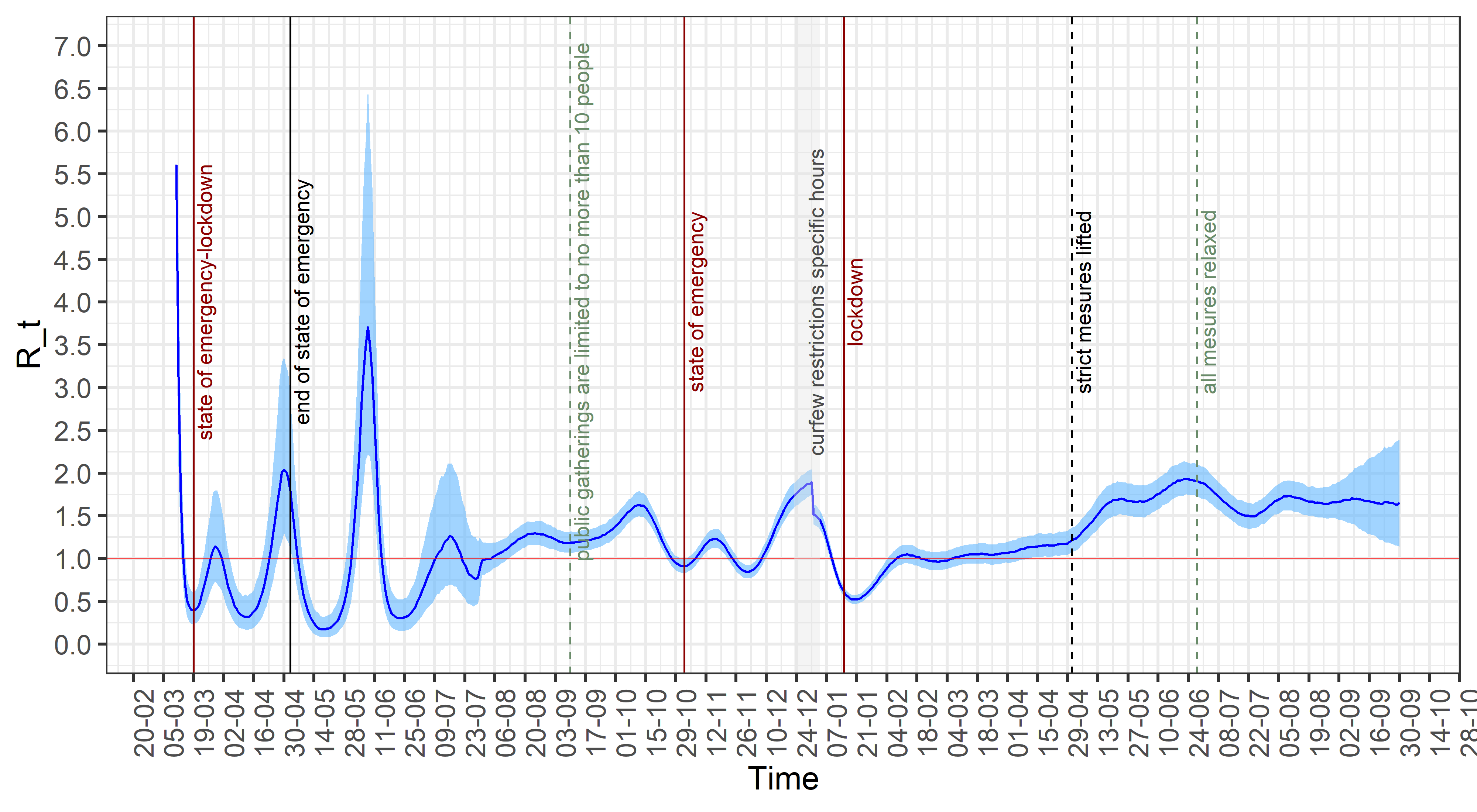}
        \end{subfigure}
\caption[Greece-Portugal - Time-varying reproduction number.]{Greece-Portugal - Time-varying reproduction number.  Medians(lines) and 50\% CI(shaded areas).}
\label{fig:R_t_GR_PT}
\end{figure}

\begin{figure}
\captionsetup{format =plain}%
     \centering
         \begin{subfigure}[b]{\linewidth}
             \centering
             \caption{\label{fig:R_UK}\textbf{United Kingdom}}
             \includegraphics[width=\linewidth]{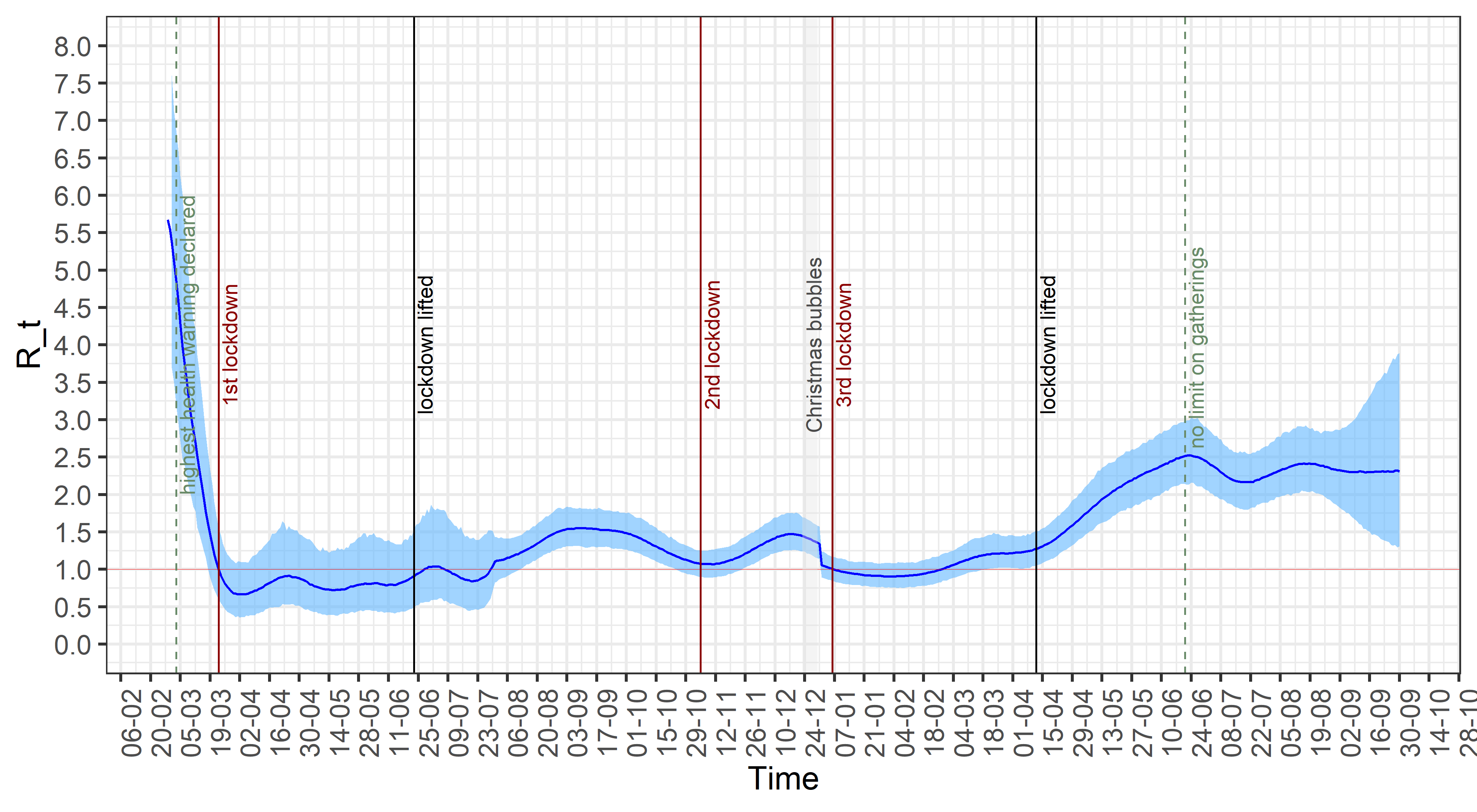}
         \end{subfigure}
         \par\bigskip
         \begin{subfigure}[b]{\linewidth}
             \centering
             \caption{\label{fig:R_DE}\textbf{Germany}}
             \includegraphics[width=\linewidth]{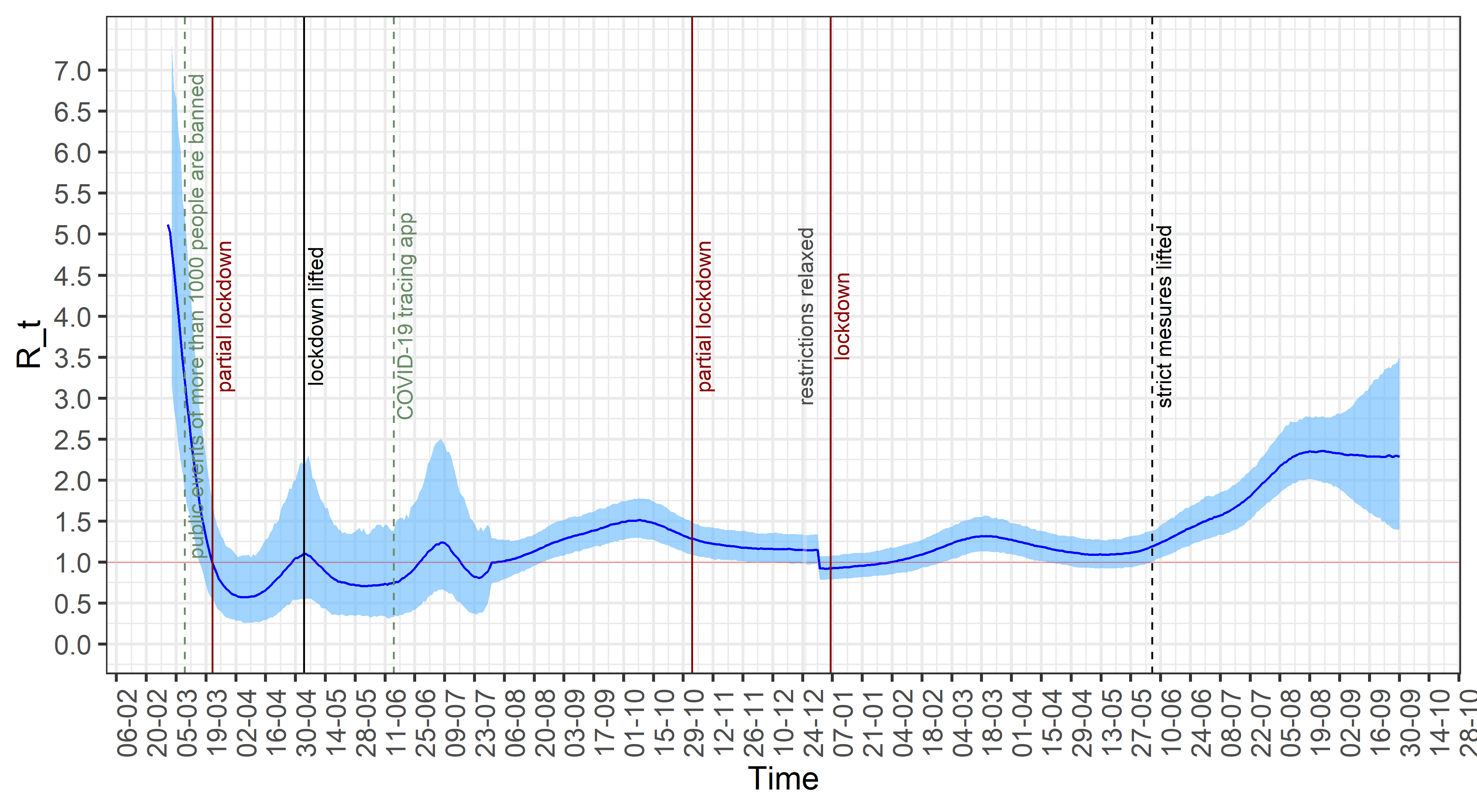}
        \end{subfigure}
\caption[United Kingdom-Germany - Time-varying reproduction number.]{United Kingdom-Germany - Time-varying reproduction number. Medians(lines) and 95\% CI(shaded areas).}
\label{fig:R_t_UK_DE}
\end{figure}

\begin{figure}
\captionsetup{format =plain}%
     \centering
         \begin{subfigure}[b]{\linewidth}
             \centering
             \caption{\label{fig:R_SE}\textbf{Sweden}}
             \includegraphics[width=\linewidth]{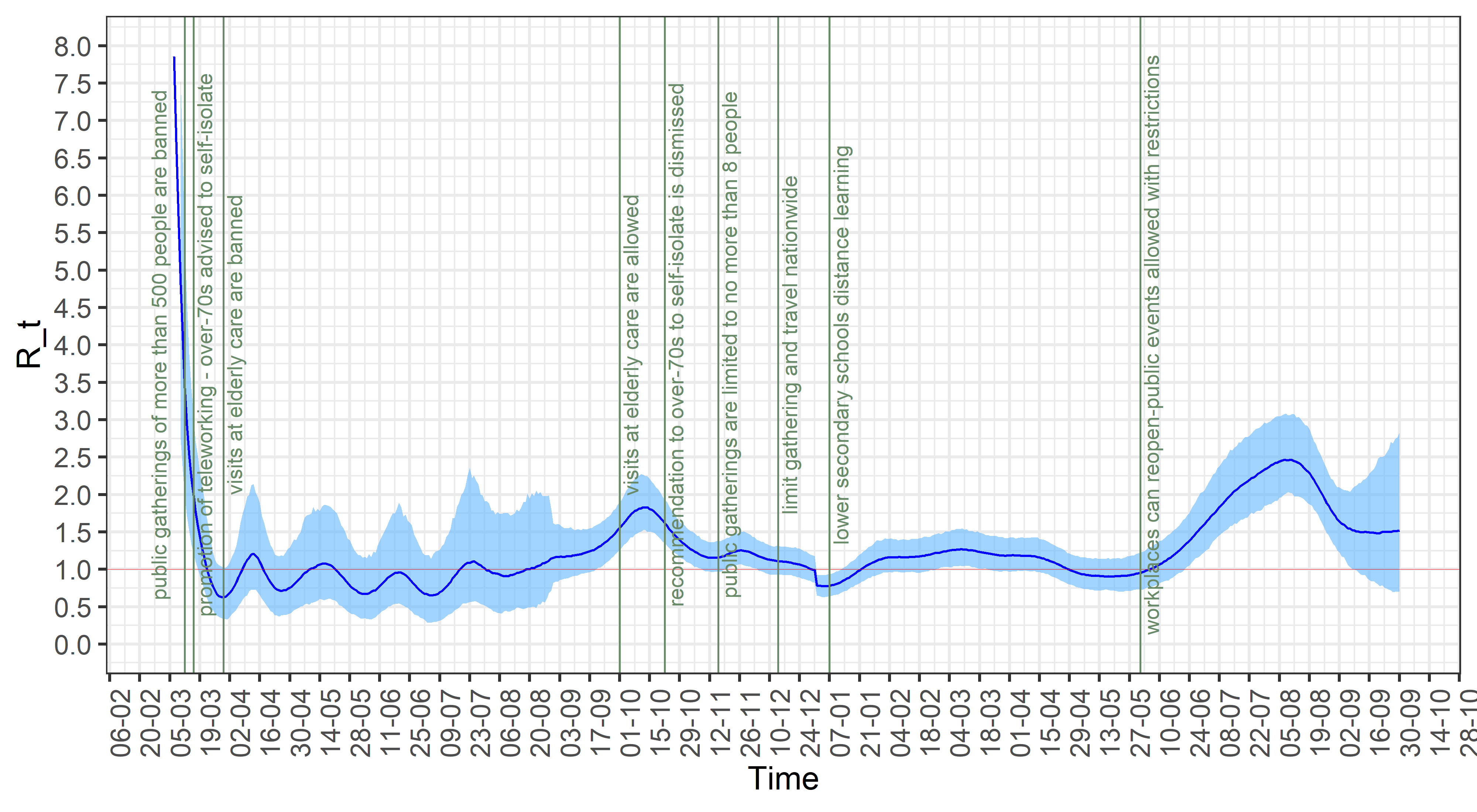}
         \end{subfigure}
         \par\bigskip
         \begin{subfigure}[b]{\linewidth}
             \centering
             \caption{\label{fig:R_NO}\textbf{Norway}}
             \includegraphics[width=\linewidth]{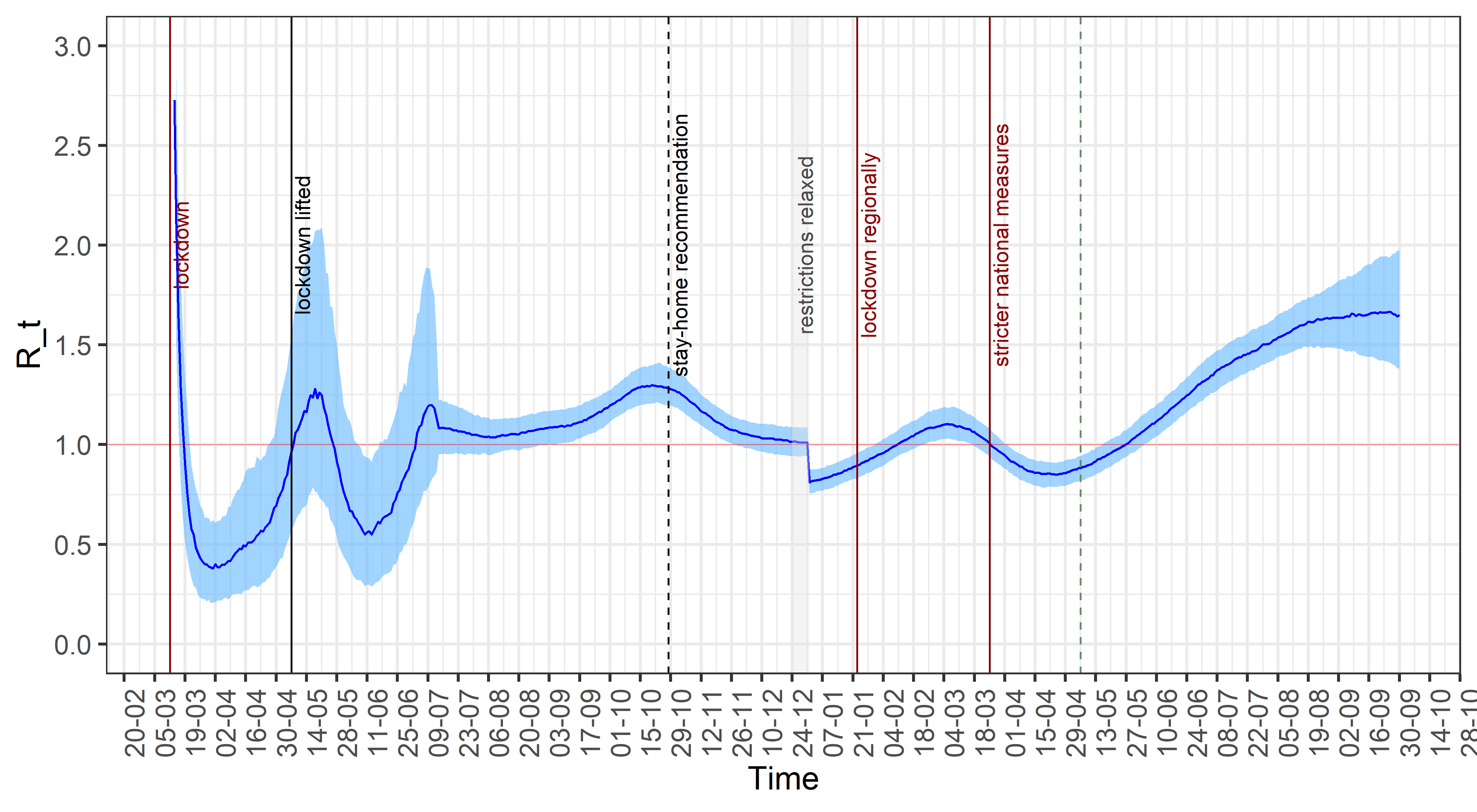}
        \end{subfigure}
\caption[Sweden-Norway - Time-varying reproduction number.]{Sweden-Norway - Time-varying reproduction number. Medians(lines) and 95\% CI(shaded areas)}
\label{fig:R_t_SE_NO}
\end{figure}

For the stochastic transmission SEIR model, the time-varying reproduction number $R_t$ can be thought of as the average number of secondary cases generated by a typical infectious individual, calculated here using $2\beta_t/\gamma_2$. Figures \ref{fig:R_t_GR_PT}, \ref{fig:R_t_UK_DE}, \ref{fig:R_t_SE_NO} summarize our results on $R_t$, for each country pair. 

Estimates of $R_t$ at the start of the epidemic must be viewed with caution since early data on deaths can not accurately reflect the transmission dynamics of local, as opposed to country-wide, spread. For $R_t$, unless otherwise stated, we report posterior medians and pointwise equal-tailed 95\% credible intervals. Analytical results for all the inferred parameters characterizing transmission, can be found in the supplementary material, including the parameters appearing in the observational process. We examine countries in pairs based on their similarities in terms of demographics and health and social welfare infrastructures. Even though our study includes only European countries, the adopted preventive measures may differ significantly between countries and so may the response of the local populations to those measures, as captured by the variation in the transmission rate. 

In early March 2020, after the detection of the first SARS-CoV-2 infections, both Greece and Portugal (Fig. \ref{fig:R_t_GR_PT}) introduced sequentially several control measures aiming to prevent large-scale outbreaks. We estimate that early measures taken by both countries, such as cancellation of large public events and closure of educational facilities, managed to drop $R_t$ significantly, before their nationwide lockdowns. In Greece, lockdown sustained a low spread of SARS-CoV-2 until the beginning of the summer, as reflected in the estimated $R_t$ which remained below 1, as opposed to Portugal for which we estimate temporary fluctuations of $R_t$ well above 1 during the same period. Between August and September, some months after the restrictive measures were eased, there was a gradual re-emergence of SARS-CoV-2 transmission in both countries. 

We estimate that the gradual reintroduction of social distancing measures led to a fall in $R_t$, before the imposition of lockdowns, which sustained lower $R_t$ in the short run. However, in early 2021, in the wake of Christmas and New Year’s relaxed measures, as well as the establishment of more transmissible variants, a steady increase in $R_t$ is estimated in Greece. As a consequence, control measures were strengthened further which resulted in lower transmission levels until June, as reflected in the estimated $R_t$. In Portugal on the other hand, even though a state of emergency was declared in early November, the transmission wasn't tamed easily, and in conjunction with relaxed measures during Christmas, we estimate that $R_t$ reached its highest level since the first wave, at the end of December 2020. Hospitals were pushed to the limit of their capacity during Portugal’s third wave which was driven by the highly transmissible alpha variant~\cite[]{davies2021estimated,volz2021assessing}. Eventually, stricter lockdown rules were imposed reducing $R_t$. As the control measures were eased during summer 2021, in the presence of the more transmissible delta variant, both countries faced significant rises in $R_t$.

Figure \ref{fig:R_t_UK_DE} illustrates the progression of estimated $R_t$ in United Kingdom and Germany. In Germany large events were cancelled and schools as well as non-essential shops were closed by the middle of March reducing $R_t$ significantly even before the partial lockdown, which however sustained $R_t$ levels well below 1. Elevated testing and tracing policy early in the outbreak, allowed Germany to start lifting restrictions in early May while maintaining low transmission. Regarding the United Kingdom, we estimate that by the middle of March the slow introduction of social distancing measures, before the nationwide lockdown, managed to reduce $R_t$ below 1. A gradual re-emergence of SARS-CoV-2 transmission after summer gave rise to $R_t$, which by mid-September is estimated to be constantly above 1 in both countries. Control measures and ultimately lockdown tamed transmission in the short-run in both countries, however, we estimate that the significant drop in $R_t$ was prominent before the implemented lockdowns. Increased transmissibility of the alpha variant as well as relaxed restrictions during Christmas, resulted in a rapid rise in infections both in the United Kingdom and Germany. Both countries implemented stricter lockdowns while the estimated $R_t$ was decreasing. Despite the stricter measures and a significant vaccine rollout in both countries, we estimate that $R_t$ remained over 1 up to the end of September in the presence of highly transmissible variants. 

While Norway’s public health response to COVID-19 was similar to that of many European countries including those under study, Sweden adopted less restrictive measures with no general lockdown. Figure \ref{fig:R_t_SE_NO} presents our results on the estimated $R_t$ for Sweden and Norway during the period under study. The day the global pandemic was declared, Norway acted quickly, imposing a lockdown with school closures and rigorous testing. As we estimate, the early measures resulted in a substantial reduction in $R_t$ well below 1. In contrast, Sweden sequentially introduced non-pharmaceutical interventions, cancelling large public events and recommending social distancing measures, especially for more vulnerable groups. However, all businesses as well as schools continued to operate. We estimate that $R_t$ managed to drop below 1, although at a smaller pace compared to Norway, but Sweden faced excess transmission in elderly care homes leading to many deaths. A resurgence of infections during the second pandemic wave led Sweden to introduce stricter control strategies, similar to those of other countries. We estimate that $R_t$ started increasing in September in both countries, however, the estimated $R_t$ in Sweden peaked at a much higher level in October, compared to Norway. Re-introduction of control measures dropped the estimated $R_t$ in the short-run for both countries, yet during the first months of 2021, increased transmissibility characterizing the third pandemic wave, led to an increase in $R_t$. Sweden faced a more severe burden with an estimated $R_t$ higher than Norway.

\subsection{Estimates of the total epidemic burden and independent validation}

The number of new daily infections is one of the main epidemiological quantities offering straightforward insight into the actual burden of the pandemic. In the supplementary material we present the estimated daily new cases, the reported cases for each country as well as the estimated aggregate incidence from March 2020 to September 2021. Our findings indicate that in all countries, during the first and second pandemic waves, the estimated daily new cases are significantly higher than the laboratory-confirmed ones. 

\begin{figure}
\centering
\captionsetup{format =plain}%
\includegraphics[width=\linewidth]{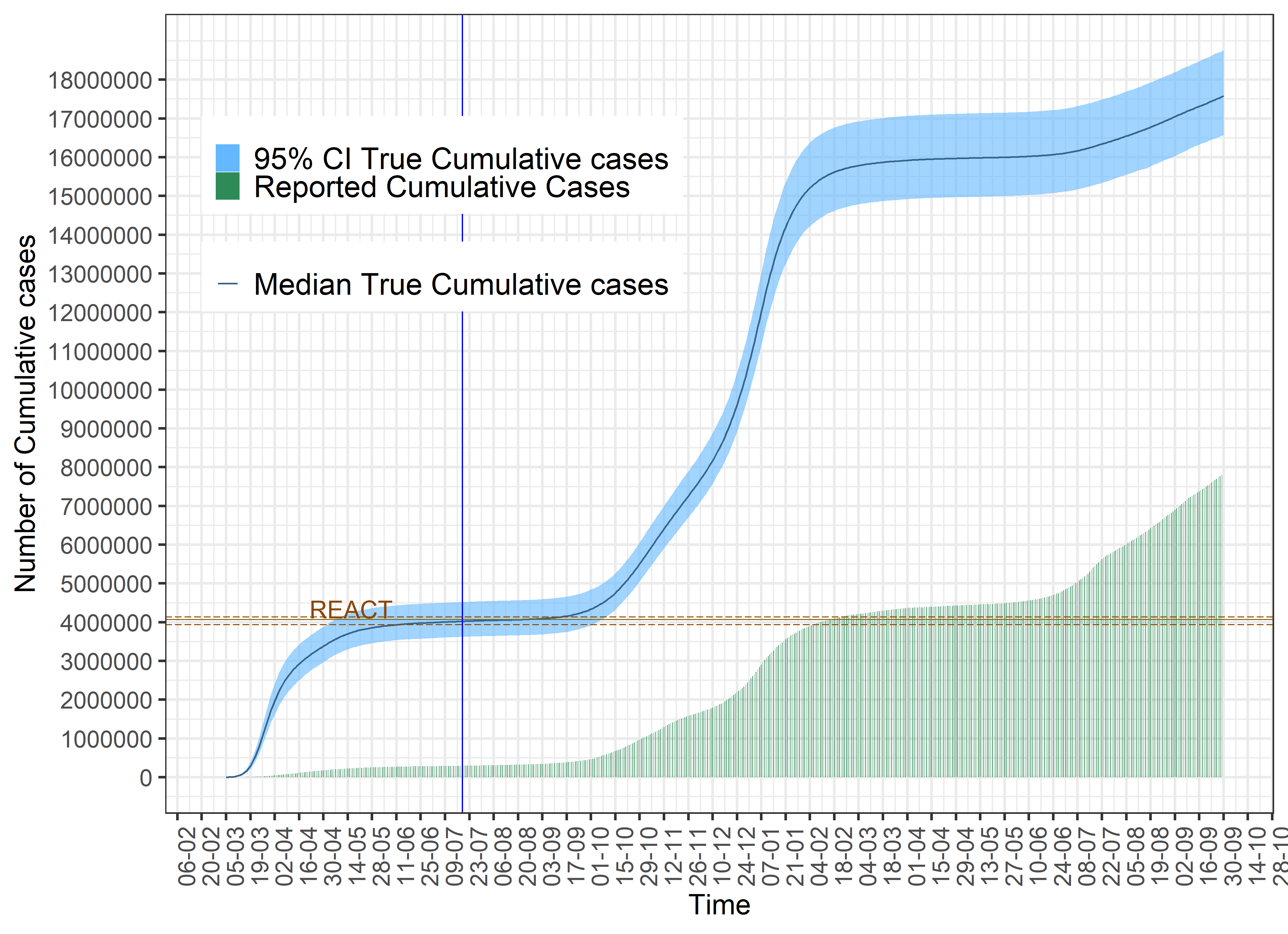}
\caption[United Kingdom - Total population infected.]{\textbf{United Kingdom} - Total population infected, data(bars), median(line) and 95\% CI(shaded area).}
\label{fig:UK_C}
\end{figure}

In Figure \ref{fig:UK_C}, the estimated number of cumulative cases in the United Kingdom is presented, along with the equivalent estimate from REACT-2~\cite[]{ward2021sars}. Seroprevalence surveys such as REACT constitute a direct approach to estimate the actual number of individuals that have been infected but have not been detected by surveillance systems. We selected this particular survey since it was carefully conducted, it was sufficiently large to offer an accurate estimate of the proportion infected and took place essentially after the first wave, therefore minimising the chance of missing cases due to waning antibody levels.

 According to REACT-2, by mid-July the overall antibody prevalence in England was $6\%$ ($95\%$ CI: $5.8-6.1$). If we adjust the estimated overall prevalence to the population in the United Kingdom, our estimates on the total population infected essentially coincide with the estimates from REACT-2, independently validating our findings.

\subsection{Estimates of additional epidemiological quantities}

Our estimates on the number of true daily cases may be combined with data on daily laboratory-confirmed cases resulting in an estimate of the number of unreported cases each day and consequently an estimate of the daily reporting ratio. Thus, using the posterior median of the estimated total number of infections at time $t$, denoted by $c_t$, we explicitly incorporate a reporting delay between actual exposure and report, $L$, so the number of unreported cases can be described as $c^{unrep}_t = c_{t-L} - c^{rep}_t$, where $c^{rep}_t$ is data on the number of laboratory-confirmed cases which are reported at time $t$ (see supplementary material). We consider a time delay between infection and report ($L$) equal to $6$ days~\cite[]{abbot2020} and define the daily reporting ratio as the ratio of laboratory-confirmed cases to the estimated total number of cases adjusted to their time of report i.e.
\begin{equation}
    r_t =c^{rep}_t/c_{t-L}
\end{equation}

\begin{figure}
     \centering
     \begin{subfigure}[b]{\linewidth}
         \begin{subfigure}[b]{0.5\linewidth}
             \centering
             \caption{Greece}
             \label{fig:GR_ratio}
             \includegraphics[width=\linewidth]{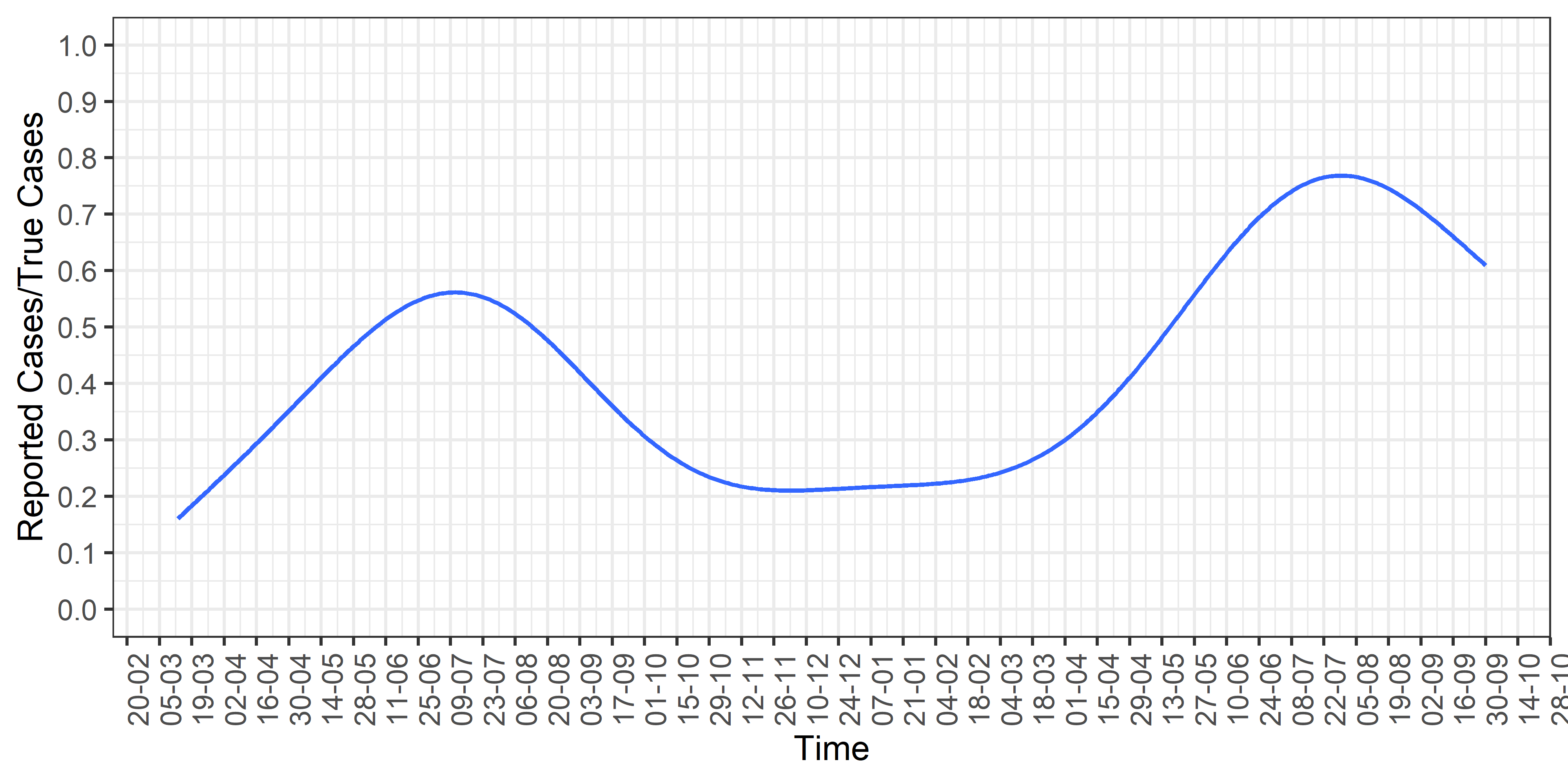}
         \end{subfigure}
         \hfill
         \begin{subfigure}[b]{0.5\linewidth}
             \centering
             \caption{Portugal}
             \label{fig:PT_ratio}
             \includegraphics[width=\linewidth]{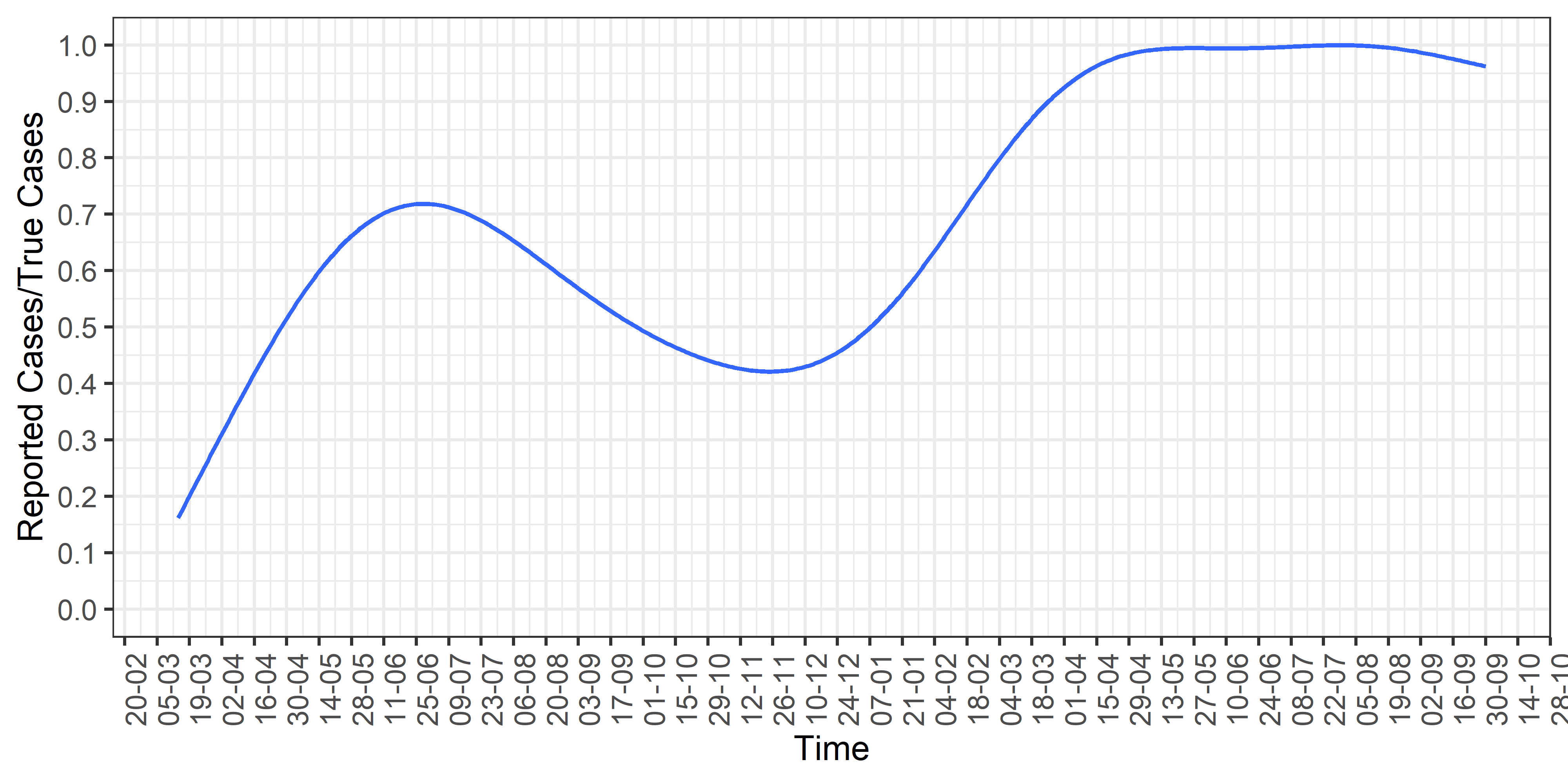}
        \end{subfigure}
     \end{subfigure}
     \hfill
     \begin{subfigure}[b]{\linewidth}
         \begin{subfigure}[b]{0.5\linewidth}
             \centering
             \caption{United Kingdom}
             \label{fig:UK_ratio}
             \includegraphics[width=\linewidth]{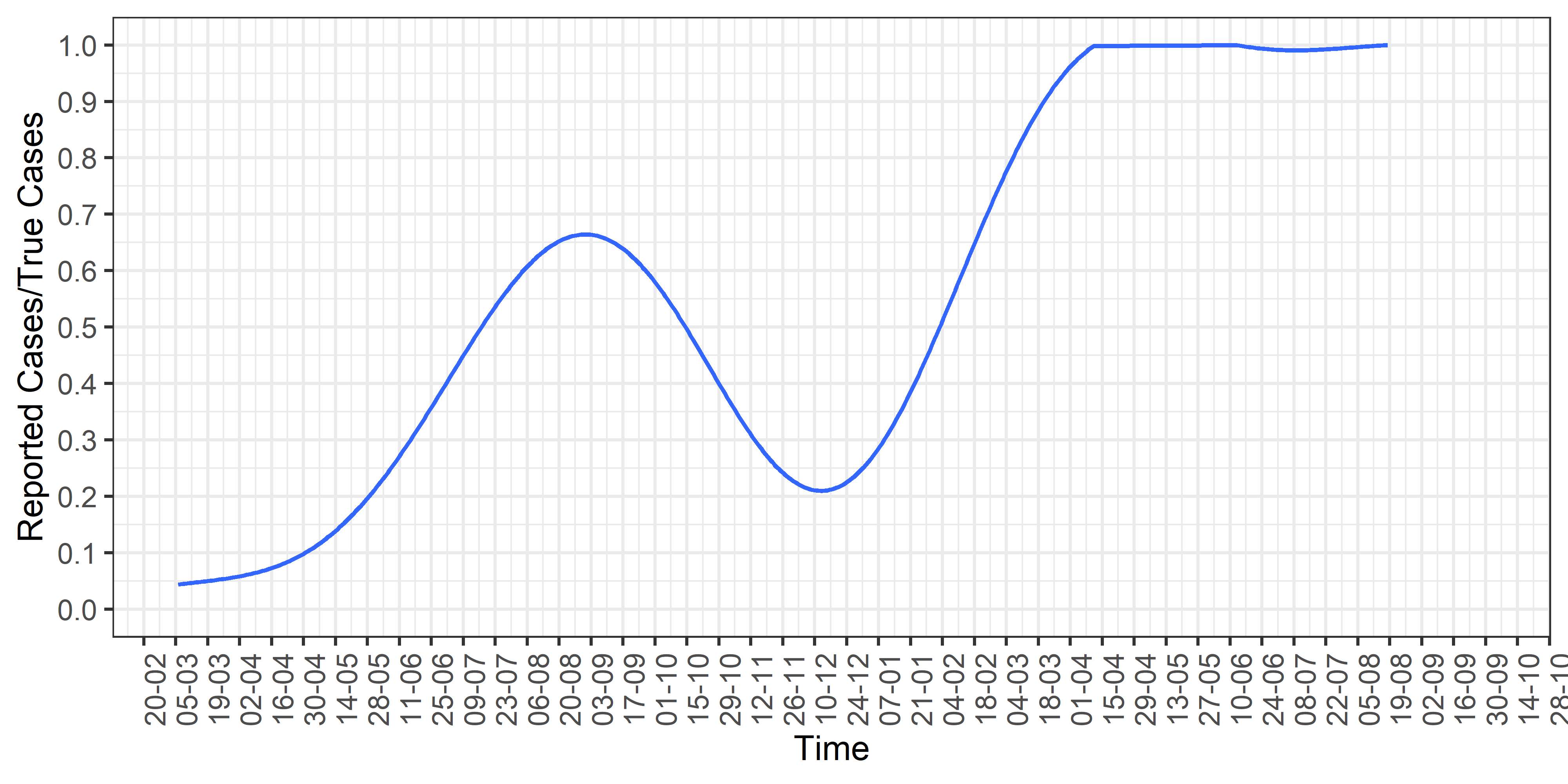}
         \end{subfigure}
         \hfill
         \begin{subfigure}[b]{0.5\linewidth}
             \centering
             \caption{Germany}
             \label{fig:DE_ratio}
            \includegraphics[width=\linewidth]{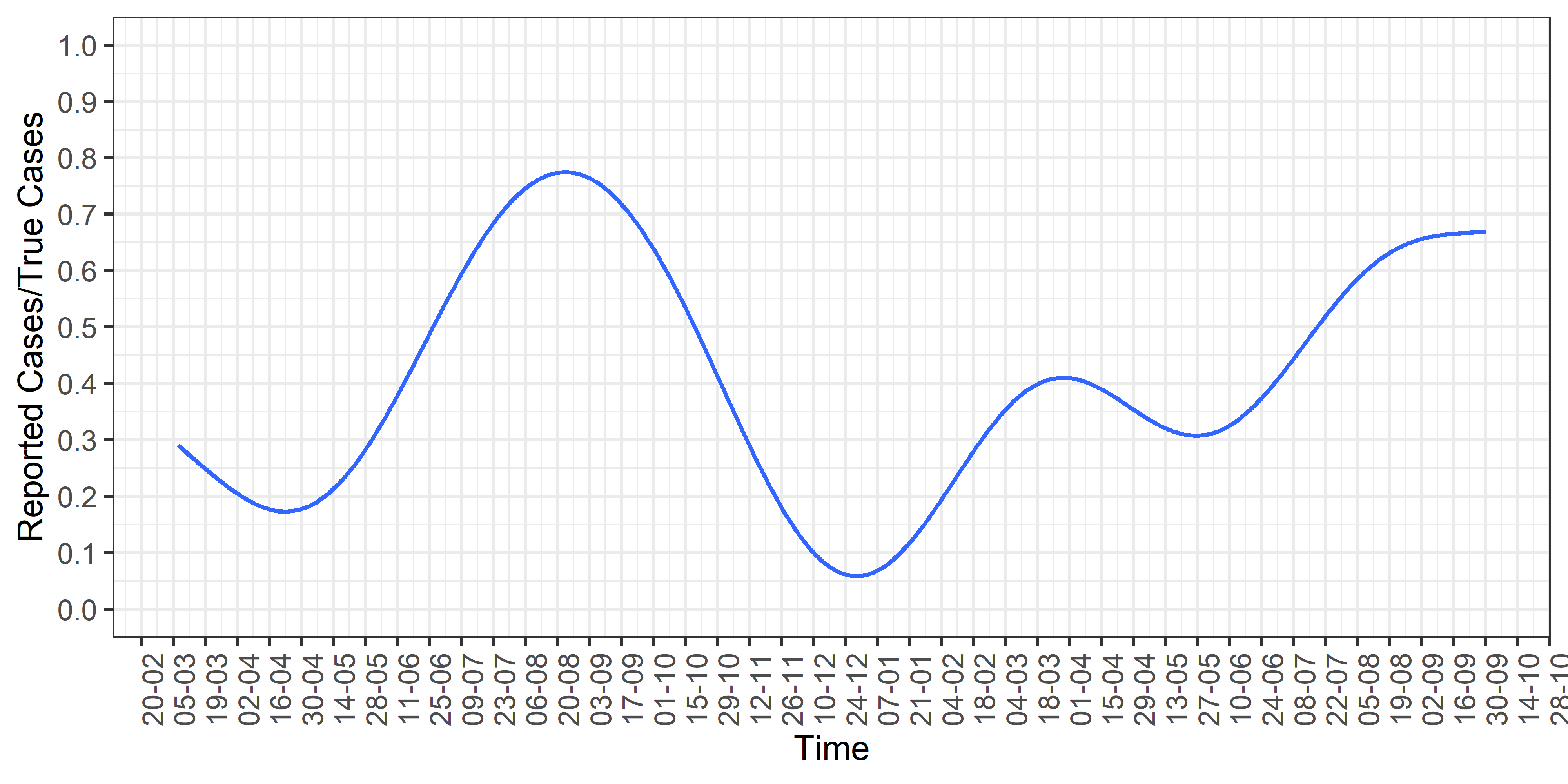}
        \end{subfigure}
     \end{subfigure}
     \hfill
     \begin{subfigure}[b]{\linewidth}
         \begin{subfigure}[b]{0.5\linewidth}
             \centering
             \caption{Sweden}
             \label{fig:SE_ratio}
            \includegraphics[width=\linewidth]{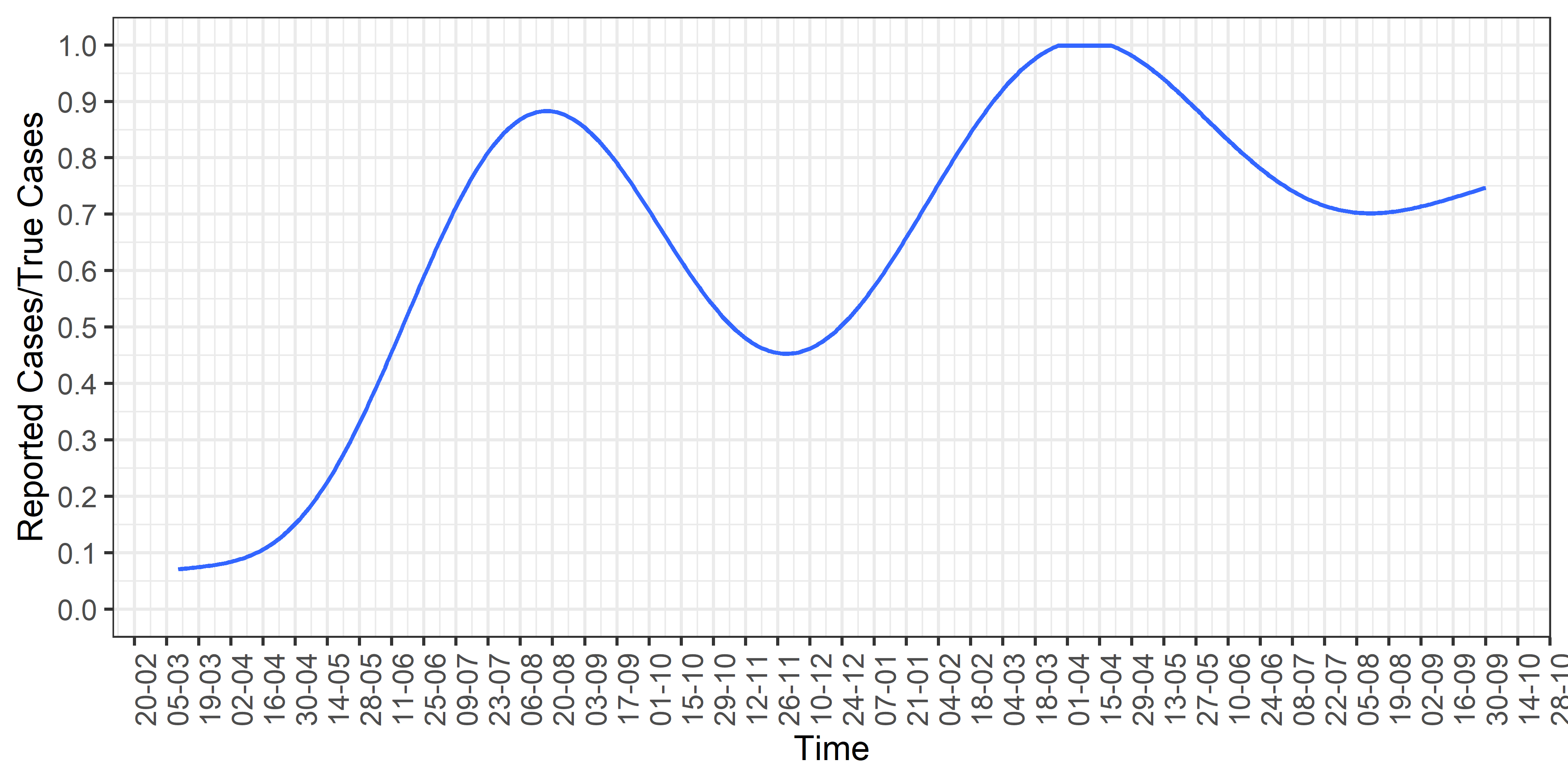}
         \end{subfigure}
         \hfill
         \begin{subfigure}[b]{0.5\linewidth}
             \centering
             \caption{Norway}
             \label{fig:NO_ratio}
            \includegraphics[width=\linewidth]{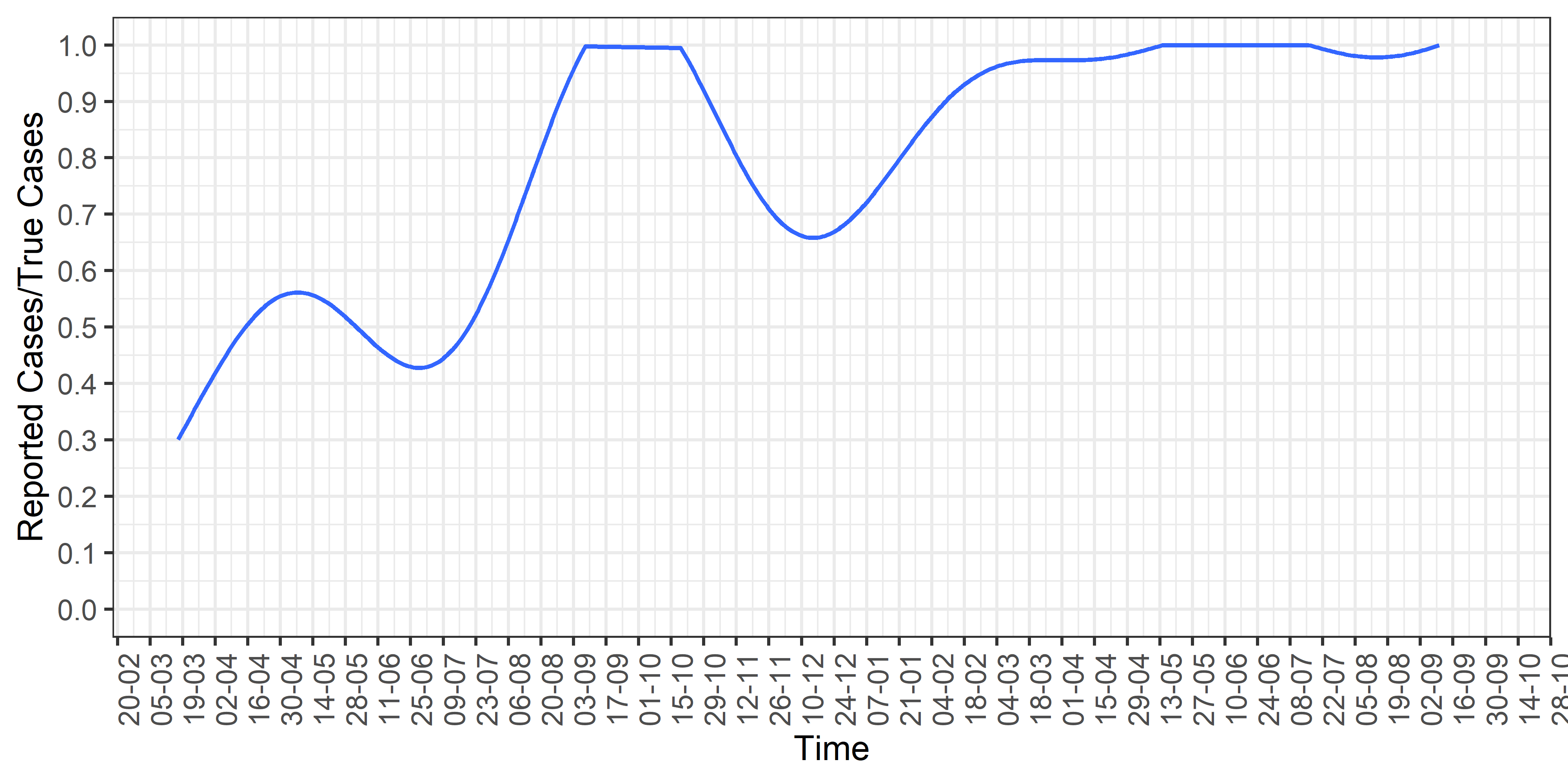}
        \end{subfigure}
     \end{subfigure}
    \caption{\label{fig:ratio}Smoothed daily reporting ratio.}
\end{figure}

Especially during the first pandemic wave, when reporting protocols had not been established, there are several days where the reported number of cases display spikes that do not represent an actual increase in cases in this particular day but inconsistencies in reporting. Given that we want to capture the general direction of the varying reporting ratio, we implement a  generalized additive model smoothing to remove those resulting peaks~\cite[]{wood2016smoothing}. We used the mgcv-package~\cite[]{wood2015package} and derive a spline-based smoother as a function of time. The results on the smoothed reporting ratio for each country are presented in Figure \ref{fig:ratio}. Our findings indicate that during periods of high transmission, there is a large proportion of under-reported infections in all countries, consistent with advice on stay at home unless needed. Even though most countries improved their testing coverage during the second and third wave, the predominance of more transmissible variants led to an increased number of infections, a large proportion of which was not detected by health systems, resulting in estimated reporting ratios lower than 40\%. Discordance between our estimates on cases and the reported data for Sweden and Norway suggests that further refinement is required in these estimates.

\subsection{Multivariate Regression Analysis}

As noted in subsection \ref{key_epi_par}, the estimated $R_t$ varies over time, which seems consistent with the expected changes as a result of the implemented control measures for COVID-19 in the countries under study. Essentially, the time-varying transmission rate $\beta_t$ captures transmission dynamics as shaped by the evolution of susceptible and infected individuals which in turn is affected by several time-varying factors such as non-pharmaceutical interventions, variations of human behaviour based on sociodemographic characteristics and climatic variations among others. These effects are highly interdependent making it difficult, in the absence of additional data sources, to disentangle their individual contributions to $\beta_t$ and design targeted and more efficient public health control strategies for each country. However, using publicly available data, we can investigate the link between some generic measures of disease control and key epidemiological quantities. In what follows we examine the relationship between the time-varying transmission rate, the daily number of true cases and the daily reporting ratio with two disease measures; mobility patterns and testing policies. Given the significant impact of vaccinations on transmission dynamics, we have chosen to run this analysis until March 2021, when the effect of non-pharmaceutical interventions was still dominant.

Social distancing measures such as stay-at-home recommendations, limitations on gatherings, closure of schools and workplaces and general restrictions on internal movement can be reflected in changes in mobility patterns~\cite[]{kraemer2020effect, lemaitre2020assessing}. These kinds of measures aim to limit the contact rate between individuals and therefore reduce transmission. Mobility data by Google~\cite[]{googlemobility} are collected by geographical location and summarize relative changes in movement in different categories of places, such as retail and recreation, groceries and pharmacies, parks, transit stations, workplaces, and residential. We access Google mobility data for the studied countries and kept the first component of a principal component analysis as a single measure of mobility~\cite[]{james2013introduction}. 

Regarding testing and tracing policies, a crude measure of the efforts to increase testing capacity in each country is the reported number of daily tests. The objective of scaling up testing policies is to tame transmission by the early isolation of confirmed or suspected cases. However, there are many different technologies for COVID-19 testing and reporting procedures differ between countries and across time. Publicly available datasets containing data on the number of tests per day and per country, are maintained in the data portal Our World in Data~\cite[]{owidcoronavirus} based on various sources and contain either both PCR and antigen tests or only PCR tests. We access these data for Greece, Portugal, Sweden, Norway and United Kingdom while for Germany we use data reported by Robert Koch Institute, Federal Ministry of Health (see supplementary material). Data of this kind may not be an accurate representation of each country's testing and tracing policy but may reasonably account for a significant fraction of the actual tests performed and crudely reflect possible variations in the testing policy.

We assume that the estimated time-varying transmission rate, true cases and reporting ratio correlate over time and use multivariate regression analysis in order to examine their relationship to mobility patterns and testing policies. We use as covariates the sum of lagged mobility trends weighted by the time they are able to generate infections and the daily number of tests performed 3 to 6 days ago. So we run the following multivariate regression for each country:
\begin{equation}
\mathbf{Y_t} \sim \mathrm{MVN}  \Biggl( \mathbf{\delta_1} m_t + \mathbf{\delta_2} tests_{t-3} + \mathbf{\delta_3} tests_{t-4}  + 
\mathbf{\delta_4} tests_{t-5} +  \mathbf{\delta_5} tests_{t-6}, \boldsymbol{\Sigma} \Biggr)
\end{equation} 
where $\mathbf{Y_t}  = \left(c_{t}, log(\beta_t), logit\left(r_t\right)\right)$. The mobility proxy $m_t$ is described as $\sum_{\tau=1}^{t-1} mob_{\tau} \pi_{t-\tau}$, where $mob$ is the first principal component of movement trends and $\pi \sim \Gamma(2.6, 0.4)$ is the serial interval which is discretized by $\pi_1 = \int_{0}^{1.5} \pi(\tau) d\tau$ and $\pi_s = \int_{s-0.5}^{s+0.5} \pi(\tau) d\tau$ for $s=2,3,...$. The number of tests at day $t$ is denoted by $tests_t$ and the regression coefficients by $\delta_i$, $i=1,\dots,5$.

The covariance matrix is represented by $\boldsymbol{\Sigma}$ and we can rewrite it in terms of the correlation matrix $\Omega$ as, $\boldsymbol{\Sigma} = D_{\sigma} \Omega D_{\sigma}$, where $D_{\sigma}$ denotes a diagonal matrix with diagonal elements $\sigma_i$, $i=1,2,3$. Then, we specify an  LKJ onion method correlation matrix distribution~\cite[]{lewandowski2009generating} for $\Omega$. The parameterization of LKJ distribution used in Stan allows to sample matrices depending on a shape parameter. The shape parameter determines whether we expect to sample posterior matrices close to the identity matrix or to general positive definite matrices~\cite[]{Stan2018Stan}. In principle, large values of the shape parameter pertain to correlations close to zero while values less than 1 suggest high probability for non-zero correlations. We use Stan's implicit parameterization of the LKJ correlation matrix in terms of its Cholesky factor. We account for the dependence in our variables by assuming that $\Omega \sim \mathrm{LkjCholesky}(0.5)$ which translates to a U-shaped prior over random correlation matrices and assign a half-Normal prior on the standard deviations: $\sigma_{i} \sim N^+(0, 10)$. 

For the vector of regression coefficients $\boldsymbol{\delta}$ we use Zellner's g-prior, a multivariate normal density with covariance matrix proportional to the inverse Fisher information matrix~\cite[]{zellner1986assessing}, i.e.
\begin{equation}
    \boldsymbol{\delta} \sim \mathrm{MVN}(0, g\sigma_{\delta}^2(X'X)^{-1})
\end{equation}
where g reflects the amount of information available in the data relative to the prior. We set $g = n$ which is equivalent to the prior having the same amount of information as one observation~\cite[]{kass1995reference}. Note that since we use the posterior medians of the responses we are likely to underestimate the uncertainty of the regression coefficients but the key findings may not be severely affected.

Our results suggest that mobility has a significantly positive effect on the transmission rate for Greece, Portugal, United Kingdom and Germany, indicating that increased mobility is associated with increased transmission. This effect is weaker with respect to the total infections as a significant effect is clear only for Portugal. The relation between the reporting ratio and mobility is not immediately apparent since increased mobility can increase infections but if at the same time the number of tests increases then the effect on the reporting ratio would be unclear. In Greece, United Kingdom, Germany, Sweden and Norway our estimates reflect a positive effect of mobility on the reporting ratio while there is a negative effect in Portugal. 

Concerning the impact of tests on transmission and infections, we would expect increased testing capacity to decrease the number of infections, as well as the transmission rate, given that confirmed infections are detected earlier and isolated. Our findings are in line with the latter, indicating a significant negative effect of the lagged number of tests on transmission rate for Greece, Portugal, Sweden and Norway. However, an increase in the tests performed may be the result of increased transmission in the community. A positive statistically significant relation between the estimated daily cases and the number of tests performed during the previous days is observed in all countries. Finally, the reporting ratio is positively associated with test numbers in the United Kingdom and Norway. In any case, our results for Sweden and Norway should be cautiously interpreted, given the discrepancies in our estimates of the daily infections.

\section{Discussion}
\label{Section4}

In this paper, we present a Bayesian approach for the estimation of temporal changes in the reproduction number of SARS-CoV-2 through data on deaths. Using a flexible stochastic extension of the SEIR model, we examine the COVID-19 pandemic in 6 European countries, inferring key epidemiological quantities such as the case reproduction number and the daily number of cases. The adopted COVID-19 outbreak control measures primarily aim to affect the transmission rate affecting the evolution of susceptible and infected populations. We assume that the generation of infections is described by an extension of the deterministic SEIR compartmental model where the transmission rate is stochastic. A proportion of these infections result to the deaths that we observe, according to a certain probability. 

We estimate that during the time course of the pandemic there have been substantially more infections than those detected by health care systems. Especially during the peak of each pandemic wave, the actual number of infections is significantly larger. The estimated cumulative cases can offer a measure of the actual burden of the pandemic. We show that the estimated changes in the reproduction number are consistent with the expected variation in SARS-CoV-2 transmission over time, as a result of the implemented control strategies. We estimate that all countries except Sweden, having introduced several non-pharmaceutical interventions, were able to drop $R_t$ below 1 well short of their nationwide lockdowns. The effects of sequentially introduced interventions in a small period of time, are highly interdependent making it hard to disentangle their individual contribution. Therefore, one may not offer robust conclusions concerning optimal strategies in the absence of additional data sources.

A distinct characteristic of our modelling approach is the absence of strong structural assumptions for the temporal evolution of the transmission rate, and subsequently for the case reproduction number, departing from the piecewise constant assumption of \citet[]{flaxman2020estimating}. Changes in $R_t$ are only driven by variations in the observed data on deaths for each country. We consider that control measures, different public responses to these measures based on cultural characteristics, adaptive human behaviour during a pandemic and any other time-varying factor are reflected in the trends in the numbers of deaths resulting from the respective infections. Therefore, our framework can be adapted to other countries in a straightforward manner.

Several limitations need to be considered when using death counts as the main source of information. Early in the pandemic, in the absence of European and international standards, some deaths due to COVID-19 may not be recorded leading to underestimation of infections. Therefore our initial estimates must be viewed with caution. Also, reporting procedures differ between countries both in terms of the timing of the report as well as the definition of COVID-19 related death. We used an integrated data source, comparing the data reported by each national health authority to the data maintained by CSSE at JHU. In addition, data on deaths depend on past infections and are not suitable for real-time analysis without further assumptions. However, in the absence of large seroprevalence studies in many countries, death counts offer a credible option for evaluating the actual burden of the pandemic in terms of people infected. 

The objective of this work was to provide a flexible framework offering an accurate representation of what has happened in the pandemic thus far. Extending our analysis for 19 months increased the computational cost and introduced several time-dependent factors which should be taken into account. Our findings rely on estimates of the $ifr$ which have large uncertainty especially after partial immunity is induced through vaccination. We allow only for deterministic changes in $ifr$ at specific time points based on our empirical observations on the emergence of more lethal and transmissible variants, the introduction of vaccines and the improvement or extreme pressure on health infrastructures. Seroprevalence studies may offer additional insights into the time-varying $ifr$. Unfortunately, such surveys are not readily available for all countries. During the time course of the pandemic factors such as the capacity of hospitals during periods of high transmission and the efficacy of the available vaccines may affect the infection to death distribution, assumed constant throughout this analysis. Additional detailed data on hospitalization can relax this assumption, further refining the results. Nevertheless, as evident by the comparison of our results to independent studies, the generic approach of this work offers a flexible and accurate framework for modelling SARS-CoV2 transmission. A natural extension of the model presented, is to deviate from the homogeneous mixing assumption, including multiple types of individuals. One such extension is presented in ~\citet[]{bouranis2022bayesian} and an alternative approach is the subject of current research. 
\newpage
\section*{Acknowledgments}
Anastasia Chatzilena states that: "Part of this research is co-financed by Greece and the European Union (European Social Fund- ESF) through the Operational Programme «Human Resources Development, Education and Lifelong Learning» in the context of the project “Strengthening Human Resources Research Potential via Doctorate Research” (MIS-5000432), implemented by the State Scholarships Foundation (IKY)". This work was part of first author's PhD thesis submitted in the Department of Economics at Athens University of Economics and Business.

\bibliographystyle{rss}
\bibliography{references}
\end{document}